# Mesostructured composite materials with electrically tunable upconversion properties

**Haridas Mundoor[1], and Ivan I Smalyukh[1,2,3]** *

[1]Department of Physics and Department of Electrical, Computer, and Energy Engineering

[2]Materials Science and Engineering Program, and Liquid Crystal Materials Research Center, University of Colorado, Boulder, Colorado, United States 80309.

[3]Renewable and Sustainable Energy Institute, National Renewable Energy Laboratory and University of Colorado, Boulder, Colorado 80309, United States.

E-mail: ivan.smalyukh@colorado.edu

**Abstract**

A promising approach of designing mesostructured materials with novel physical behavior is to combine unique optical and electronic properties of solid nanoparticles with long-range ordering and facile response of soft matter to weak external stimuli. Here we design, practically realize, and characterize orientationally ordered nematic liquid crystalline dispersions of rod-like upconversion nanoparticles. Boundary conditions on particle surfaces, defined through surface functionalization, promote spontaneous unidirectional self-alignment of the dispersed rod-like nanoparticles, mechanically coupled to the molecular ordering direction of the thermotropic nematic liquid crystal host. As host is electrically switched at low voltages ≈ 1V, nanorods rotate, yielding tunable upconversion and polarized luminescence properties of the composite. We characterize spectral and polarization dependencies, explain them through invoking models of electrical switching and upconversion dependence on crystalline matrices of nanorods, and discuss potential practical uses.

## 1. Introduction

Conversion of infrared to visible light is of interest for potential applications in solar energy harvesting, biological imaging, laser designs, high-capacity storage devices, etc. Although one can achieve this with frequency conversion using nonlinear optical materials, [1,2] the requirement of high-intensity coherent infrared light sources in this approach limits the usage of such process in many applications. An alternative method involves an upconversion mechanism based on sequential absorption of two or more infrared photons of lower energy to yield a single photon of higher energy in the visible spectral region. [3-6] Owing to the fact that this process can be initiated by low-intensity, incoherent light, upconversion stands out as a promising process for applications involving infrared-visible energy transformation,[7-9] unrestricted to uses of coherent infrared light sources. Typically, doping of solid crystals with rare earth ions ($RE^{3+}$), such as Erbium ($Er^{3+}$) and Thulium ($Tm^{3+}$), is used to mediate the upconversion processes, which are enabled by presence of a large number of excited metastable states. Although low-absorption cross-sections of such ions limit efficiency of the corresponding up-conversion processes, one can overcome this limitation by further co-doping the host solid matrix with sensitizer ions like ytterbium ($Yb^{3+}$). The larger absorption cross-section of the sensitizer ions mediates increased up-conversion efficiency enabled by transferring photon energy to the luminescent centers. This enhanced infrared-visible transformation can be implemented in upconversion nanoparticles (UCNPs) of variable shape and size that can be synthesized in large quantities [10-12]. In addition to potential uses in materials science, UCNPs with lanthanides doping have been identified as promising alternatives for deep-tissue biomedical tracking and imaging due to their unique properties, such as the narrow-line emission, and nonblinking emission,[13-15] which can be tuned over a wide spectral range by selecting suitable dopant ions. Although the spectral features of the lanthanides ions remain largely unaffected by the host materials, the upconversion efficiency depends on the solid matrix structure of the ion-hosting medium [16-18] and often leads to polarized emission of individual nanoparticles due to the intra-ions transition properties and local crystal symmetry of rare earth ions. [19] However, such polarization-dependent properties lack tunability by external stimuli and are not inherited by colloidal nanoparticle dispersions due to random orientations of these anisotropic particles within dispersions, indicating the requisites of a host materials which can direct and control the location and orientations of nanoparticles. A promising approach of designing such mesostructured materials with novel physical behavior is to combine unique optical and electronic properties of solid nanoparticles with long-range ordering and facile response of soft matter to weak external stimuli. Liquid crystals (LCs) [20] recently emerged as host media mediating novel types of colloidal self-assembly and self-alignment. [21-25] For example, high-concentration bulk dispersions of anisotropic nanometer-sized metal particles in thermotropic LCs can be aligned either parallel or perpendicular to the LC director ***n*** by suitably modifying surface functionalization of the particles, enabling low-voltage switching of their orientation.[24-25]

In this work, we report orientationally ordered dispersion of rod-shaped UCNP particles in a nematic LC medium, which can be controlled by defining surface boundary conditions for the LC director at the confining cell surfaces and reversibly switched at low voltages. The ordered dispersions of UCNP particles in nematic LC medium were achieved through polymer surface-functionalizing the constituent solid particles, providing tangential orientation of the LC molecules at the UCNP surfaces. These surface boundary conditions mediate anisotropic UCNP-LC interactions and spontaneously align the particles along ***n*** through the free energy minimization of the nematic host. We study novel polarization-dependent, electrically switchable optical properties of the UCNP-LC mesostructured composite and explain their origin by invoking simple models based on long-range nanoparticle alignment induced by the host medium. LC switching and polarization dependent properties of the UCNP-LC composite are correlated with the crystal structure of the host matrix of the rod-like semiconductor nanoparticles. Finally, we discuss these findings in the context of general prospects for achieving novel physical behavior of mesostructured organic-inorganic composites by integrating facile response of soft matter systems, such as LCs, with unique optical properties of solid nanostructures.

## 2. Results and Discussions

We have used three types of particles in our experiments: $\beta$-NaYF$_4$ particles with 30%Gd$^{3+}$, 18%Yb$^{3+}$, and 2%Tm$^{3+}$ (referred to as "UCNP1" particles from now on), $\beta$-NaYF$_4$ particles with 60%Gd$^{3+}$, 18%Yb$^{3+}$, 2%Tm$^{3+}$ (UCNP2), and $\beta$-NaYF$_4$ particles with 30%Gd$^{3+}$, 18%Yb$^{3+}$, and 2%Er$^{3+}$ (UCNP3). Concentration of Gd$^{3+}$ ions plays an important role in defining the growth and crystallinity of NaYF$_4$, yielding a hexagonal atomic crystal lattice for the particles. [11] The crystalline structure of NaYF$_4$ induces polarization anisotropy of UCNP emission, which is arising from the crystal symmetry.[19] **Figure 1** represents the energy level diagram based on the energy values for $Er^{3+}$ and $Tm^{3+}$ ions obtained from the previous literature reports [6,26-29]. The crystalline phase of the host matrix also induces fine-structure splitting of the energy levels due to crystal field interaction [26-29]. Synthesis of β-$NaYF_4$ UCNPs doped with $Yb^{3+}$, $Tm^{3+}$ and $Gd^{3+}$ yield rod-shaped particles with rather uniform size distribution, such as the ones shown in the SEM images in **Figure 2a**. For dispersing in 4-Cyano-4-pentylbiphinyl (5CB), the original oleic acid (OA) capping of the as-synthesized particles is replaced with Methoxy poly ethylene glycol silane (Si-PEG), thereby providing weak tangential boundary conditions for the LC molecules with weaker distortion of the LC director around the particles, as schematically shown in Figure 2b. Thus, similar to plasmonic gold nanorods with PEG capping,[24] on-average, orientations of the dispersed rod-like UCNPs follow *n*, although orientations of individual particles are influenced by thermal fluctuations, which tend to deviate their orientations away from this free-energy-minimizing direction, so that the scalar orientational order parameter is smaller than unity (Figure 2c). Luminescence microscopy of the low-concentration samples, filled in a planar aligned cell of 30 μm thickness, reveals good dispersion of UCNPs in the LC medium, showing Brownian motion of bright spots of size comparable to diffraction-limited optical resolution and corresponding to single particles (Figure 2d). The dispersions of UCNP particles were found stable and retaining both orientational ordering and average inter-particle distances for at least four days if kept in the nematic phase. At higher concentrations, these particles are at separations too close to be resolved in diffraction-limited optical microscope observations. Low-magnification microscopy images of the LC composite shows uniform brightness in the cell region, very different from the dark background at the cell edge without such particles (Figure 2d, inset). Luminescence microscopy is also used as a means to check the concentrations of the particles dispersed in the LC medium, which we find consistent with values based on sample preparation. Further, the particles are also visible in the dark field microscopy images, an example of which is shown in Figure 2e, although UCNPs have relatively low polarization-dependent refractive contrast relative to the LC medium (refractive index of the particles is ≈ 1.6, in-between the values of ordinary ≈ 1.5 and extraordinary ≈ 1.7 indices of 5CB). Polarizing optical microscopy images (Figure 2f,g) reveal the planar alignment of the UCNP-LC composite of quality similar to that commonly achieved for pristine LCs and metal nanoparticle dispersions in LCs. [24] Alignment of the far-field LC director parallel to one of the crossed polarizers (Figure 2g) provides a means to detect possible distortions of the director field induced by individual nanoparticles. The obtained images indicate that such elastic distortions of the LC alignment around nanoparticles are minimal, so that the LC matrix is barely perturbed by the incorporated UCNPs, as depicted in Figure 2b,c. This is consistent with our findings for metal nanoparticles, [24] although UCNPs are somewhat larger in size as compared to nanoparticles used in our previous studies and may induce slightly stronger elastic distortions as compared to their metal counterparts.

To study three-dimensional ordering and optical properties of UCNP dispersions in LCs, we probe the polarization dependent emission of the composite using excitation light provided by a 980 nm, linearly polarized output from the Ti:Sapphire laser. The used power of the excitation laser beam was limited to 10 mW to avoid laser-induced realignment, heating and ensure that the LC medium is in a nematic phase, ruling out a possibility of aggregate formations due to an unintended local transition to isotropic phase. **Figure 3a** shows emission spectra collected from the planar-aligned UCNP-LC composite containing UCNP1 particles. The detected emission corresponds to the transition $^3H_4$ - $^3H_6$ of $Tm^{3+}$ ions and is partially polarized along the axis of the rods (Figure 3a), consistent with previous studies for individual nanoparticles. [19] We also observed the emission lines representing the transitions between other energy levels at 450, 475, 650 and 695 nm; however, these lines are weaker because of the relatively low excitation power used in our experiments. [19] We observed that the intensities of various emission peak remains unaffected by the polarization state of the excitation beam, hence we fix polarization of the excitation beam to be parallel to ***n*** for subsequent measurements. However when the emission intensity is recorded by rotating the analyzer (**A**) inserted right before the spectrometer, the emission shows maximum intensity for the $^3H_4 - ^3H_6$ transition for the analyzer **A** ∥ ***n***, gradually decreasing when the analyzer is rotated away from this orientation, with a minimum intensity observed for **A** ⊥ ***n*** (Figure 3a). The peak emission intensity follows the expected ∝ $\cos^2\theta$ dependence, as shown in the inset of Figure 3a. Experiments with the UCNP2 dispersions yield qualitatively similar results, as summarized in Figure 3b.

The facile response of the LC host to external electric field translates to the similar response of anisotropic particles dispersed in this medium, since the particle orientations are mechanically coupled to the LC director ***n*** through surface anchoring boundary conditions for ***n***. Our results show that the UCNPs spontaneously orient parallel to ***n*** at no external fields and then follow it during electrical switching. Figure 3c shows the schematic representation of response of the LC director $n$ across the cell depth and the corresponding response of dispersed particles. Figure 3a,b shows the luminescence spectra changes of UCNP-LC composite in response to the electric field (20V, 1KHz) for UCNP1 and UCNP2 particles dispersed in the LC. We observed maximum field-induced changes in the spectral intensity for the initial orientation of the analyzer **A** ∥ **n** at no field, corresponding to the re-alignment of **n** and particles from a planar-aligned to predominantly perpendicular configuration with respect to the glass substrates and **A** (Figure 3c). However in the initial **A** ⊥ **n** case, the orientations of **n** and particles with respect to **A** remain unaffected, as expected, consistent with the relatively negligible voltage-induced spectral intensity variation (Figure 3a,b). We have also characterized voltage dependence of luminescence light intensity and polarized emission characteristics of the UCNP-LC composite during switching (Figure 3d). The threshold-like behavior is observed for peak luminescence intensity versus voltage dependencies, with a sharp change of luminescence above a certain realignment threshold voltage comparable to that of pristine LCs and in dilute dispersions of gold nanoparticles in LCs [24] (Figure 3d). Although the observed spectral variation of UCNP-LC composite are similar for UCNP1 and UCNP2 particles, the difference in their luminescence characteristics could be observed by probing the luminescence polarization index η, defined as η=($I_0$- $I_U$)/($I_0$ +$I_U$), as shown in the inset of Figure 3d. Here $I_0$ and $I_U$ represent intensities of the emission peak before and after applying switching voltage $U$ across the cell. The observed polarization index is higher for the UCNP-LC composite prepared with UCNP2 particles, which is expected because of the stronger polarization anisotropy of UCNP2 particles due to highly crystalline hexagonal order induced by the relatively higher concentration of $Gd^{3+}$ ions. [30]

In order to study the polarization dependent properties of $Er^{3+}$-doped particles, we have dispersed UCNP3s in 5CB and filled in planar aligned cells with a cell gap spacing of 30 μm. The emission efficiency was found to be independent of the polarization state of excitation light relative to orientations of nanorods along **n**, an expected behavior consistent with that of single nanoparticles of similar kind.[19] However, luminescence of the UCNPs dispersed in the LC composite is partially polarized (Figures 3 and 4). **Figure 4a** shows emission spectra of the UCNP-LC composite, with a transition between energy levels of $Er^{3+}$ ions representing $^4F_{9/2}$ to $^4I_{15/2}$ and corresponding to red emission part of the spectra, as well as $^4S_{3/2}$ - $^4I_{15/2}$ and $^4H_{11/2}$ - $^4I_{15/2}$ transitions representing the green emission from the particles. These transitions are schematically represented using an energy diagram shown on Figure 1b. In order to probe the polarization anisotropy of the UCNP- LC composite, we further analyzed the emission spectra by rotating the orientation of **A** with respect to the LC director ***n***. The comparison of emission spectra obtained at **A** ∥ ***n*** and **A** ⊥ ***n*** reveals salient features of these spectra (Figure 4). We find higher intensity for the lower-wavelength spectral part of the red emission (655 nm) at **A** ⊥ ***n***, lower intensity of emission at higher-wavelength part of green emission (552 nm) and lower intensity corresponding to the shoulder peak at 525 nm for **A** ⊥ ***n***. We have characterized the variation of the emission peaks by plotting relative changes of intensity of the emission spectra at 655 nm, 552 nm and 525 nm versus the orientation of **A**, as shown in Figures 4b,c,d, respectively. The angular intensity variations follow ∝ $\cos^2\theta$ or ∝ $\sin^2\theta$ dependencies on the angle $\theta$ between **A** and ***n***, with mutually orthogonal intensity variations observed for the peak intensity at 655 nm (∝ $\sin^2\theta$) relative to the peaks at 552 nm and 525 nm (∝ $\cos^2\theta$), respectively, similar to such observations in previous studies of individual nanoparticles, where this behavior was linked to the symmetry of the transition dipoles leading to the emission lines.[19] The emission spectra show similar changes in the spectra for **A** ∥ ***n*** when electric field is applied across the cell, but there is only very minimal variation of luminescence intensity with voltage in the **A** ⊥ ***n*** case (Figure 4e). The dependencies of peak intensity on voltage are threshold-like, as shown in Figure 4f using examples of UCNP1 and UCNP2 particles. Although inclusion of UCNPs can slightly affect the LC properties, the observed ***n***-realignment threshold of the UCNP-LC is of the order of 1V, comparable to that of pristine 5CB. For pristine LCs confined in cells with planar geometry, similar to the one used in our studies, the threshold voltage $U_{th} = \pi[K_{11}/(\varepsilon_0\Delta\varepsilon)]^{1/2}$ is the voltage at which ***n*** starts to re-orient in response to the field due to the electric torque (tending to align ***n*** along the applied field) overcoming the combination of elastic and surface anchoring torques that tend to keep the LC undistorted in the initial planar geometry, where $\Delta\varepsilon$ is dielectric anisotropy ($\Delta\varepsilon = 8.2$ for 5CB), $\varepsilon_0 = 8.854\times10^{-12}$ C V$^{-1}$ m$^{-1}$ and $K_{11}$ is the splay elastic constant ($K_{11} = 5.4$ pN for pristine 5CB). Depending on the volume fraction of UCNPs in the composite, $U_{th}$ was found to be 1.2-1.9 times larger than that of pure 5CB (Figures 3 and 4),[24] which can be attributed to the role of incorporated particles in decreasing the effective $\Delta\varepsilon$ and increasing elastic constants of the ensuing composite medium as compared to pure LC.

The observed changes in the particles emission spectra mediated by reorientation of nanoparticles can be understood considering the fine structure splitting of the energy levels, induced by crystal field and local site symmetry of crystalline matrix at $Er^{3+}$ locations. The energy levels responsible for the luminescence properties of the $Er^{3+}$ ions result from the spin orbit interaction between the unpaired electrons, which are degenerate with respect to the spin quantum number. [6] However when the ions are located in a crystal lattice of the solid nanoparticles, this degeneracy is lifted due to the crystal field interaction resulting in the splitting of energy levels, [26-29] yielding fine energy level structures ($E_{cr}$) schematically shown in Figure 1b. The strength of these crystal field mediated transitions depends on the symmetry of the transition dipoles, with respect to the axes of the local coordination polyhedron.[26-29] The UCNP3 particles have predominantly hexagonal crystal lattice due to the presence of $Gd^{3+}$ ions,[11] although pure hexagonal phases are observed only at higher concentrations of $Gd^{3+}$. [30] Moreover presence of $Gd^{3+}$ ions in $NaYF_4$ crystal, magnify the local crystal asymmetry by altering the bond angle and bond length between $RE^{3+}$ and $F^-$ .[19] We believe that the observed polarization effects of UCNP3 particles are related to the crystal field induced transition levels, affected by local crystal symmetry of $Er^{3+}$ ions. Although, crystal field induced spectral lines could not be resolved in our room temperature spectroscopy measurements, with our low resolution spectrometer, the variation of intensities at the edges of the luminescence spectra (552 nm and 655 nm) with rotation of **A** indicates the variation of relative intensities of these lines. However, the residual polycrystallinity present in the UCNPs limits the polarization anisotropy of the individual particles. [30] This limitation known for individual particles translates to a similar constraint of the studied composite with orientationally ordered nanorods (Figure 3 and 4). Tunability of the emission properties of the UCNP-LC composite can be enhanced further by optimizing the structural and geometrical properties of the UCNP particles, in particular, by making nanoparticles more monocrystalline while still having desired ion doping and composition. We also note that, unlike for LC dispersions of other types of nanorods with polarization-sensitive optical characteristics,[24,25] polarized emission data like the ones presented in Figure 3 and 4 cannot be used to determine the scalar orientational order parameter of UCNPs due to the effects of partial polycrystallinity, which would be possible for monocrystalline UCNPs. Although, different hydrothermal synthesis method could be used to obtain single crystalline, rod shaped UCNP particles,[12] larger particle size obtained with such method limits the concentration of UCNP in 5CB due to the larger distortion of LC director caused by the particles.

Physical underpinnings behind the spontaneous alignment of rod-like UCNPs in a nematic host matrix are similar to the case of metal nanorod dispersions capped with similar polymeric ligands.[24] Since the PEG provides weak tangential surface boundary conditions for *n*, surface anchoring free energy is minimized for nanorod orientation along *n* of the host LC. However, UCNPs are somewhat larger in size as compared to, for example, gold nanorods we studied previously,[24] so that the spontaneous alignment of such particles in 5CB may be additionally mediated by the minimization of the elastic free energy due to week elastic distortions around the inclusions, which for rod-like particles with tangential boundary conditions also favors alignment along *n*. [31-33] In addition to the polymer capping layers, the stability of our colloidal dispersions may be also enhanced by screened electrostatic interactions due to residual charged groups on particle surfaces originating from the synthesis process. It is known that nematic colloidal particles can be charged [34] and that the screening of inter-particle electrostatic interactions by counterions dissolved in the non-polar [35] thermotropic nematic host like 5CB can yield relatively long-range repulsive interactions that can be responsible for relatively large inter-particle distances at varying volume fractions of UCNPs in 5CB (**Figure 5**). Since many UCNP surface functionalization approaches have been developed in designing their uses for biomedical applications, it may be possible to reliably control the type and strength of surface boundary conditions as well as the strength of electrostatic and elastic interactions to obtain different forms of orientational and positional ordering of UCNPs in LCs, which will be a goal of our future studies.

To study stability of UCNPs dispersed in LCs, we have analyzed inter-particle spacing for the three concentrations of UCNPs. Figure 5 represents distributions of the time-averaged center-to-center interparticle spacing, measured based on luminescence microscopy images collected from the respective samples. We have tracked hundreds of particles at different locations of the sample. The lateral spacing between the nearest neighbors was measured by observing these particles using video microscopy over tens of seconds. We found Gaussian distributions, with the mean inter-particle spacing shifted towards lower separation values with increasing particle concentration. The mean inter-particle spacing estimated from the distribution curves are found to be 1.25 μm, 0.8 μm and 0.47 μm for UCNP concentrations 0.85 pmol/ml, 3.24 pmol/ml and 16 pmol/ml respectively. The inter-particle separations at the highest concentration samples (16 pmol/ml and higher) are smaller than the optical resolution of our microscope, so that the individual particles within the dispersions could not be resolved using luminescence and dark field imaging approaches. It is interesting that stable UCNP dispersions in nematic LC hosts can be obtained up to such high concentrations, indicating interesting possibilities for new forms of soft matter organization and switching to be explored in future studies. Although beyond the scope of our present study, it is important to note that the distributions of inter-particle distances at different concentrations indicate presence of relatively long-range repulsive inter-particle forces, which could have electrostatic origin and exhibit long-ranged repulsions due to the non-polar nature of the nematic LC host.

Since LCs allow for dispersing multiple types of colloidal inclusions, the optical properties of UCNP-LC composites can be tuned further by co-dispersing other optically responsive colloidal inclusions like plasmonic particles with UCNP particles, which could lead to potential applications in energy conversion, multicolor displays etc. Since we have recently shown that the upconversion efficiencies of the individual UCNP particles can be modified with plasmonic structures, [36] the orientation and ordering of the metal particles with respect to the UCNP particles could be used to tune the composites for obtaining a desired optical response. Proper designing of the composite system with suitable inclusions of optically responsive entities may give rise to new composite properties, much different from the individual particles, with diverse applications. The upconversion efficiency of UCNPs in nematic hosts can be also enhanced by using the core-shell geometry of particles.[37] Since the upconversion properties of UCNPs are effectively independent of the background refractive index of the host LC medium but surface plasmon resonance properties of metallic nanoparticles exhibit a strong dependence on it, co-assembly and alignment of metal and semiconductor nanoparticles may thus provide a new means of designing novel composites that inherit unique properties of metal and semiconductor nanoparticles as well as the response of host soft matter medium to weak external stimuli, as needed for many applications. Our approach can be also extended to LC elastomers, where it may provide means of generating visible and ultraviolet light needed for photo-actuation by using infrared excitation that can penetrate deep into such often strongly scattering media. From a fundamental standpoint of view, colloidal dispersions of nanoparticles in LCs may lead to formation of new mesophases with unusual combinations of orientational and partial positional ordering in the molecular host medium and colloids themselves. Combined with pre-engineered director structures, the modification of polarization of luminescent light of UCNPs in various LC hosts, such as cholesterics, may be used to obtain novel sources of light, potentially including distributed-feedback mirror-free cholesteric lasers.

## 3. Conclusion

In conclusion, we have demonstrated dispersion, long-range orientational ordering, and electrical switching of anisotropic upconversion nanoparticles in a nematic LC. These UCNP-LC composites inherit unique optical properties of individual UCNP nanoparticles and facile response and ordering of the LC host matrix at the same time, yielding tunable upconversion and polarization dependent luminescence properties. We explained our findings considering LC switching that causes rotation of rod-like particles while following the LC host reorientation, as well as the crystal symmetry of $NaYF_4$. More generally, our study showed that doping at different levels, ranging from doping of crystal lattice of solid nanoparticles by atomic ions to doping of LC host medium by these nanoparticles could be used synergistically in designing new composite materials with properties of interest controlled by weak external stimuli.

## 4. Experimental Section

*Nanoparticles synthesis and dispersion in nematic LC:* We synthesized lanthanide-doped β-$NaYF_4$ nanorods capped with OA by following the general synthesis method recently reported in literature.[11] The synthesis procedure was utilizing hydrothermal process, carried out under high temperature and pressure inside a teflon lined autoclave (Hydrion Scientific, 25 ml). The chemicals used for the experiments, Ytterbium Chloride Hexahydrate ($YbCl_3$ $6H_2O$), Yttrium Chloride Hexahydrate ($YCl_3$ $6H_2O$), Erbium Chloride Hexahydrate ($ErCl_3$ $6H_2O$), Thulium Chloride Hexahydrate ($TmCl_3$ $6H_2O$), Gadolinium Chloride Hexahydrate ($GdCl_3$ $6H_2O$), Ammonium Fluoride ($NH_4F$) and OA were purchased from Sigma Aldrich. Sodium Hydroxide (NaOH) was purchased from Alfa Aesar. Si-PEG ($M_W$ -5K) were obtained from JenKem Technology, USA. Briefly, 375 mg of NaOH was mixed with 1.875 ml of deionized (DI) water, 6.25 ml of ethanol and 6.25 ml of OA. 2.5 ml of 2 M solution of lanthanide chlorides with desired molar fraction, as needed for a specific doping concentration, was mixed with this solution. Finally, 1.25 ml of 2M solution of $NH_4F$ was added to the solution and stirred for 30 min to ensure proper mixing of the precursors. The final solution was transferred to the teflon lined autoclave, kept at an elevated temperature of 200º C for 2 h and then naturally cooled down to room temperature, completing the synthesis procedure. After the synthesis, the ensuing semiconductor nanoparticles, collected at the bottom of the reaction vessel, were removed by centrifugation, washed with ethanol and water multiple times, and finally re-dispersed in cyclohexane. In order to obtain good dispersion of the nanoparticles in a nematic LC, we capped the nanoparticles with Si-PEG through the following multi-step process. First, the OA capping layer of the nanoparticles was removed by treating them with a mild acidic solution. [38] In a typical process, 6 mg of OA- functionalized UCNPs dispersed in 8 ml of cyclohexane were mixed with 4 ml of DI water with small amount of hydrochloric acid (HCl) to yield pH ≈ 4. The solutions were subjected to magnetic stirring for 2 h; during this process, the OA molecules initially attached to the semiconductor surface of the nanoparticles became protonated and started mixing with cyclohexane similar to free unattached oleic acid molecules. At the same time, the uncapped UCNPs were transferred to DI water. Following this, the UCNP particles dispersed in DI water were separated from the solution by mixing acetone followed by centrifugation. The nanoparticles were washed several times with acetone and finally dispersed in DI water. During the next sample preparation stage, to achieve Si-PEG capping, 1 ml of ethanol solution of Si-PEG was prepared by dissolving 60 mg of Si-PEG under mild heating and added, drop-by-drop, to the 5 ml of UCNP dispersion in water. The dispersion was stirred continuously for 12 h and the Si-PEG capped UCNP particles were separated from the initial solution by centrifugation (10000 RPM), until eventually well dispersed in ethanol. We then adopted a nematic nanoparticle dispersion method described in Ref. 24 to re-disperse our semiconductor nanoparticles into 5CB, (obtained from EM Chemicals). Briefly, 50 μL of UCNP particles dispersion in ethanol were mixed with 13 μL of 5CB in a 0.5 mL centrifuge tube. In order to remove the ethanol from the medium the LC mixture was kept at 90°C for 3 h, until all the ethanol was fully evaporated. The tube was sonicated for 1 min in water at a temperature of 90°C and then cooled down. This was accompanied by vigorous stirring until the initially isotropic 5CB transitioned to the nematic phase. The mixture was then centrifuged at 3000 rpm for 5 min in order to remove any residual aggregates formed during the phase transition, yielding a uniform dispersion of individual nanoparticles. Planar LC cells were prepared using glass plates coated with transparent indium tin oxide (ITO) electrodes on their inner surfaces. These confining surfaces were rubbed to impose the unidirectional boundary conditions for *n* and then glued together with UV-curable NOA-65 glue (Norland Products, Inc.) containing 30 μm silica spacers to define the desired cell gap thickness. In order to apply electric field across the cell, wires were soldered to the ITO-coated surfaces.

*Optical microscopy and spectroscopy:* Polarizing, bright-field and dark-field optical microscopy studies were performed using an Olympus BX-51 upright polarizing optical microscope with 20× dry objective with numerical aperture NA = 0.5 and a 100× oil immersion objective with adjustable NA = 0.6 - 1.3 (all from Olympus). Optical micrograph and movie capturing was done using a charge coupled device (CCD) camera (Spot 14.2 Color Mosaic, Diagnostic Instruments, Inc.). Luminescence microscopy of UCNPs dispersed in 5CB was carried out using an Olympus IX-71 inverted optical microscope equipped with appropriate dichroic mirror and spectral filters. Nanoparticles were excited using 980 nm continuous-wave output from a diode laser focused into the sample using a 4× objective, while the UCNP-upconverted light from the sample was collected either by a 20× or a by a 100× objective, and finally sent to an electron multiplying CCD camera (EMCCD, iXon3 888, Andor Technology). Luminescence spectra of the UCNP-LC composites were measured using a spectrometer USB2000-FLG (Ocean Optics), mounted on an inverted IX-81 microscope (Olympus). The nanoparticles were excited with a 980 nm, linearly polarized, pulsed output from a Ti:Sapphire oscillator (140 fs, 80 MHz, Chameleon Ultra II, Coherent) and a low-magnification objective (4×), while the luminescence spectra were collected using a 20× (NA = 0.5) objective in the forward detection geometry. A half-wave or quarter-wave retardation plate was mounted immediately before the objective to control the polarization of the excitation laser light. The luminescence signals from the samples were passed through suitable optical filters to block the excitation line. An analyzer was introduced into the optical path immediately before the spectrometer, thereby controlling linear polarization of the luminescence signal transmitted to the spectrometer and detector.

**Acknowledgements**

This research was supported by the U.S. Department of Energy, Office of Basic Energy Sciences, Division of Materials Sciences and Engineering, under Award ER46921, contract DE-SC0010305 with the University of Colorado Boulder. We thank Paul Ackerman, Jao van de Lagemaat, Taewoo Lee, Qingkun Liu, Bohdan Senyuk, and Yuan Zhang for technical assistance and discussions at different stages of this research project. We also thank A. Sanders for his assistance with imaging and acknowledge the use of the Precision Imaging Facility at NIST, Boulder for the electron microscopy imaging (Figure 2a).

**References**

[1]  N. J. Durr, T. Larson, D. K. Smith, B. A. Korgel, K. Sokolov, A. Ben-Yakar, *Nano Lett.* **2007**, 7, 941.

[2]  K. T. Yong, J. Qian, I. Roy, H. H. Lee, E. J. Bergey, K. M. Tramposch, S. L. He, M. T. Swihart, A. Maitra, P. N. Prasad, *Nano Lett.* **2007**, 7, 761.

[3]  W. Zheng, P. Huang, D. Tu, E. Ma, H. Zhuab, X. Chen, *Chem. Soc. Rev.* **2015**, 44, 1379.

[4]  F. Wang, X. G. Liu, *Chem. Soc. Rev.* **2009**, 38, 976.

[5]  G. Chen, H. Qiu, P. N. Prasad, X. Chen, *Chem. Rev.* **2014**, 114, 5161.

[6]  F. Auzel, *Chem. Rev.* **2004**, 104, 139.

[7]  R. Scheps, *Prog. Quantum Electron.* **1996**, 20, 271.

[8]  E. Downing, L. Hesselink, J. Ralston, R. Macfarlane, *Science* **1996**, 273, 1185.

[9]  B. M. van der Ende, L. Aarts, A. Meijerink, *Phys. Chem. Chem. Phys.* **2009**, 11, 11081.

[10] C. Li, J. Yang, Z. Quan, P. Yang, D. Kong, J. Lin, *Chem. Mater.* **2007**, 19, 4933.

[11] F. Wang, Y. Han, C. S. Lim, Y. Lu, J. Wang, J. Xu, H. Chen, C. Zhang, M. Hong, X. Liu, *Nature* **2010**, 463, 1061.

[12] F. Zhang, J. Li, J. Shan, L. Xu, D. Zhao, *Chem. Eur. J.* **2009**, 15, 11010.

[13] J. Zhou, M. X. Yu, Y. Sun, X. Z. Zhang, X. J. Zhu, Z. H. Wu, D. M. Wu, F. Y. Li, *Biomaterials* **2011**, 32, 1148.

[14] J. Liu, Y. Liu, Q. Liu, C. Li, L. Sun, F. Li, *J. Am. Chem. Soc.* **2011**, 133, 15276.

[15] Y. Liu, M. Chen, T. Cao, Y. Sun, C. Li, Q. Liu, T. Yang, L. Yao, W. Feng, F. Li, *J. Am. Chem. Soc.* **2013**, 135, 9869.

[16] D. M. Yang, C. X. Li, G. G. Li, M. M. Shang, X. J. Kang, J. J. Lin, *Mater. Chem.* **2011**, 21, 5923.

[17] Q. Q. Dou, Y. Zhang, *Langmuir* **2011**, 27, 13236.

[18] Q. M. Huang, J. C. Yu, E. Ma, K. M. Lin, *J. Phys. Chem. C* **2010**, 114, 4719.

[19] J. Zhou, G. Chen, E. Wu, G. Bi, B. Wu, Y. Teng, S. Zhou, J. Qiu, *Nano Lett.* **2013**, 13, 2241.

[20]   P. M. Chaikin, T. C. Lubensky, *Principles of Condensed Matter Physics*, Cambridge Univ. Press: Cambridge, **2000**.

[21]   P. Poulin, H. Stark, T. C. Lubensky, D. A. Weitz, *Science* **1997**, 275, 1770.

[22]   C. P. Lapointe, T. G. Mason, I. I. Smalyukh, *Science* **2009**, 326, 1083.

[23]   B. Senyuk, Q. Liu, S. He, R. D. Kamien, R. B. Kusner, T. C. Lubensky, I. I. Smalyukh, *Nature* **2013**, 493, 200.

[24]   Q. Liu, Y. Yuan, I. I. Smalyukh. *Nano Lett.* **2014**, 14, 4071.

[25]   Y. Zhang, Q. Liu, H. Mundoor, Y. Yuan, I. I. Smalyukh, *ACS Nano* **2015**, 9, 3097.

[26]   J. B. Gruber, D. K. Sardar, R. M. Yow, B. Zandi, E. P. Kokanyan, *Phys. Rev. B* **2004**, 69, 195103.

[27]   J. D. Carey, *J. Phys.: Condens. Matter.* **2009**, 21, 175601.

[28]   H. Steinkemper, S. Fischer, M. Hermle, J. C. Goldschmidt, *New J. Phys.* **2013**,15, 053033.

[29]   R. Kolesov, K. Xia, R. Reuter, R. Stöhr, A. Zappe, J. Meijer, P. R. Hemmer, J. Wrachtrup, *J. Nat. Commun.* **2012**, 3, 1029.

[30]   D. T. Klier, M. U. Kumke, *J. Phys. Chem. C* **2015**, 119, 3363.

[31]   F. Brochard, P. G. De Gennes, *Journal de Physique* **1970**, 31, 691.

[32]   I. I. Smalyukh, J. Butler, J. D. Shrout, M. R. Parsek, G. C. L. Wong. *Phys. Rev. E* **2008**, 78, 030701(R).

[33]   Q. Liu, J. Tang, Y. Zhang, A. Martinez, S. Wang, S. He, T. J. White, I. I Smalyukh, *Phys. Rev. E* **2004**, 89, 052505.

[34]   S. A. Tatarkova, D. R. Burnham, A. K. Kirby, G. D. Love, E. M. Terentjev, *Phys. Rev. Lett.* **2007**, 98, 157801.

[35]   J. W. Merrill, S. K. Sainis, E. R. Dufresne, *Phys. Rev. Lett.* **2009**, 103, 138301.

[36]   Qi-C. Sun, H. Mundoor, J. C. Ribot, V. Singh, I. I. Smalyukh, P. Nagpal, *Nano lett.* **2014**, 14, 101.

[37]   W. Feng, X. J. Zhu, F. Y. Li, *NPG Asia Mater.* **2013**, 5, e75.

[38]   N. Bogdan, F. Vetrone, G. A. Ozin, J. A. Capobianco, *Nano Lett.* **2011**, 11, 835.

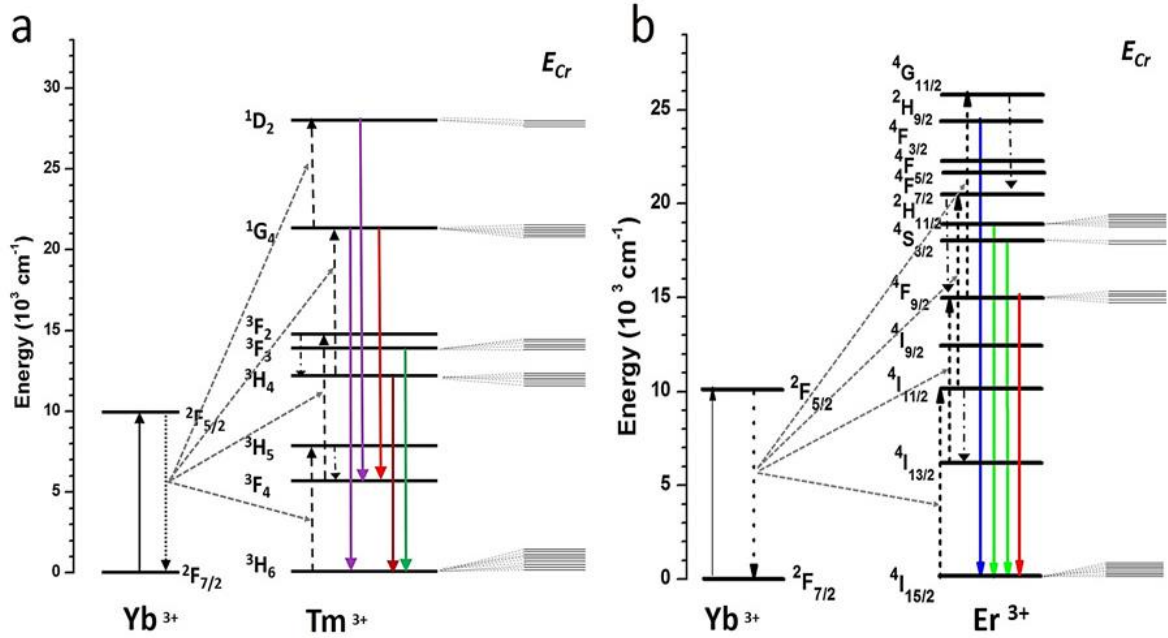

**Figure 1.** Energy level diagram for the upconversion process implemented through the energy transfer from $Yb^{3+}$ ions to $Tm^{3+}$ (a) and $Er^{3+}$ (b) ions, representing the upconversion energy transfer (dashed line) and multiphoton (dashed dot) process. The colored arrows indicate the emission lines corresponding to the transition between $^1D_2 - {}^3F_4$ (450 nm), $^1G_4 - {}^3H_6$ (475 nm), $^1G_4 - {}^3F_4$ (650 nm), $^3H_4 - {}^3H_6$ (800 nm), $^3F_3 - {}^3H_6$ (695 nm) levels of $Tm^{3+}$ ions in (a) and $^2H_{9/2} - {}^4I_{15/2}$ (410 nm), $^2H_{11/2} - {}^4I_{15/2}$ (525 nm), $^4S_{3/2} - {}^4I_{15/2}$ (550 nm), $^4F_{9/2} - {}^4I_{15/2}$ (660 nm) of $Er^{3+}$ ions in (b). $E_{cr}$ represents the energy level fine splitting due to the crystal field interaction.

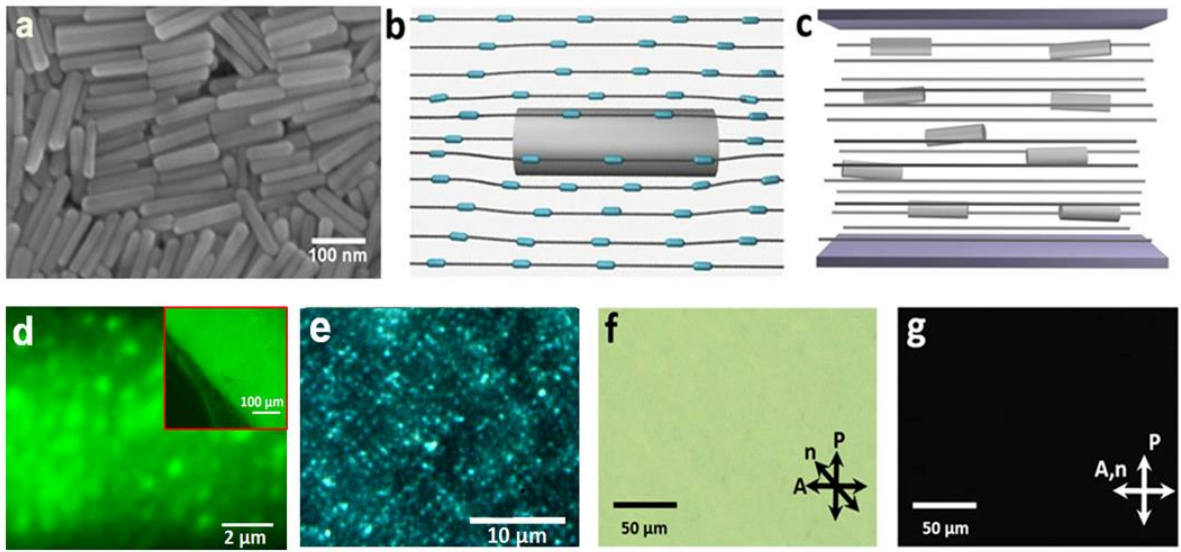

**Figure 2.** Nematic dispersions of UCNPs. a) An SEM image of UCNP1 particles drop coated on a silicon substrate. b) Schematics of a UCNP particles with weak tangential boundary conditions for rod-like LC molecules (cyan); dimensions of LC molecules ($\approx$ 1 nm in size), much larger UCNP nanoparticles, and cell gap thickness are shown not to scale. c) Schematic of dispersion of rod-like UCNP particles in LC while exhibiting a scalar order parameter lower than unity. d) High magnification luminescence microscopy image of UCNPs dispersed in 5CB. The inset depicts a low-magnification image of a high-concentration (16-20 pmol/ml) dispersion of UCNP particles in the LC, confirming uniform dispersion of UCNP particles. e) Dark-field microscopy images of UCNP particles dispersed in 5CB, revealing good dispersion quality. (f,g) Polarizing optical micrographs of the UCNP-LC composite obtained between crossed polarizer **P** and analyzer **A**, with *n* at (f) $45^0$ and (g) parallel to **A**. The estimated concentration of the particles is 0.85 pmol/ml (d,e) and 16 pmol/ml (f,g) in 5CB.

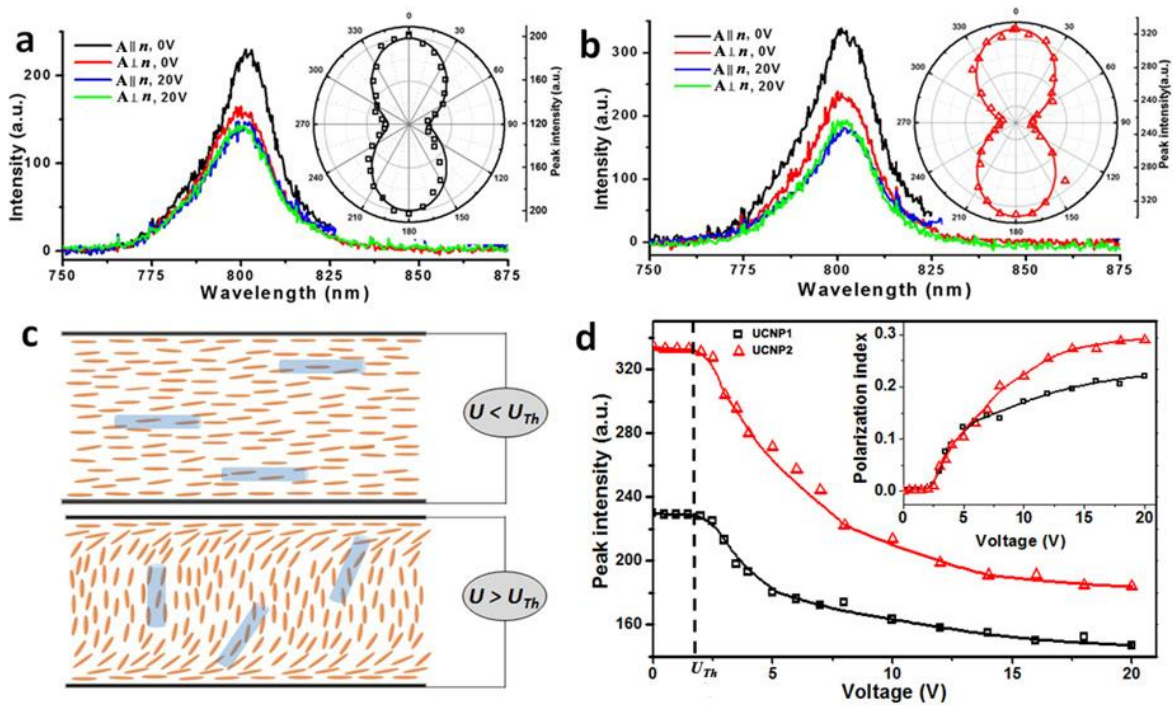

**Figure 3.** Polarization-dependent emission spectra and switching of a planar-aligned UCNP-LC composite cell. (a,b) the emission spectra for dispersions containing (a) UCNP1 and (b) UCNP2 were measured for **A** ∥ **n** (black) and **A** ⊥ **n** (red) at no external fields and at applied sinusoidal voltage U = 20V at 1kHz (blue, green), applied across the cell using transparent ITO electrodes. Insets in (a) and (b) shows polar plots representing variation of the emission peak intensity corresponding to the 0-360$^0$ rotation of **A** for the respective UCNPs. c) Schematics of the alignment of nematic LC molecules (red) and UCNPs (light blue) dispersed in the LC at no or weak applied fields (top) and at applied voltage above the realignment threshold voltage U > U$_{th}$ (bottom). Dimensions of LC molecules, UCNP nanoparticles, and cell thickness are shown not to scale. d) Variation of the peak emission intensity centered at 800 nm for UCNP1 (□) and UCNP2 (Δ) particles, for voltages ranging within U = 0 - 20 V applied across the cell, for **A** ∥ **n**. Inset shows the plots of the polarization index η for respective particles. The estimated concentration of the particles is 16 pmol/ml in 5CB.

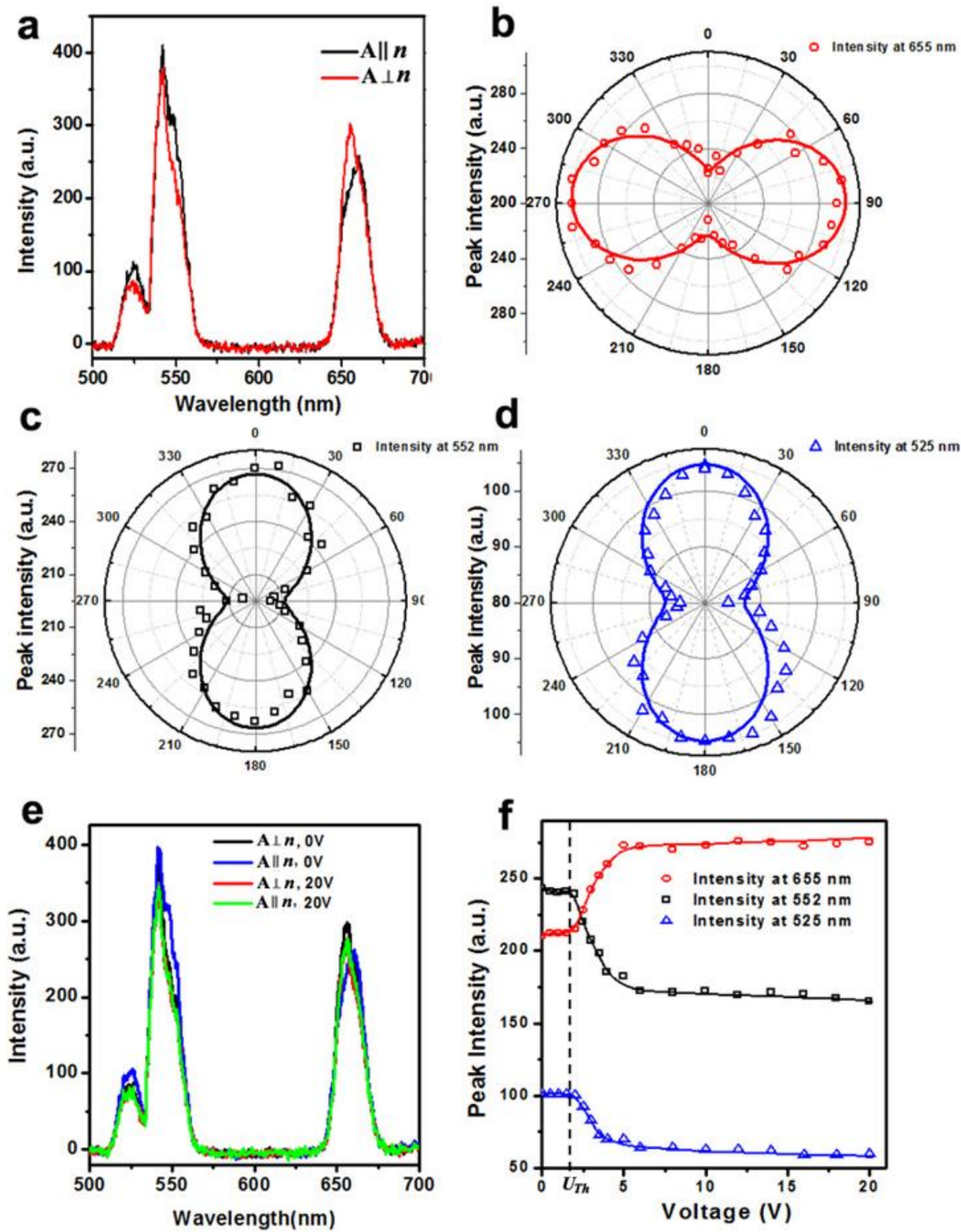

**Figure 4.** Emission spectra measured using a planar aligned UCNP-LC cell with UCNP3 particles. (a) The graphs show polarization-dependent emission spectra for **A** ∥ **n** (black) and **A** ⊥ **n** (red).

(b,c,d) Polar plots showing relative changes of the emission intensity at 655 nm (○), 552 nm (□) and 525 nm (Δ) corresponding to the 0-360⁰ rotation of **A** with respect to ***n***, as marked on the corresponding figure parts. (e) Voltage-induced changes of emission spectra at 20V, 1kHz, obtained for **A** ∥ ***n*** (green) and **A** ⊥ ***n*** (red). Similar spectra at no applied voltage are represented by black and blue curves. f) Variation of intensity versus voltage at 655 nm (○), 552 nm (□) and 525 nm (Δ), obtained for voltages varied within U = 0 - 20V and for the analyzer orientation **A** ∥ ***n***. The estimated concentration of the particle is 16 pmol/ml in 5CB.

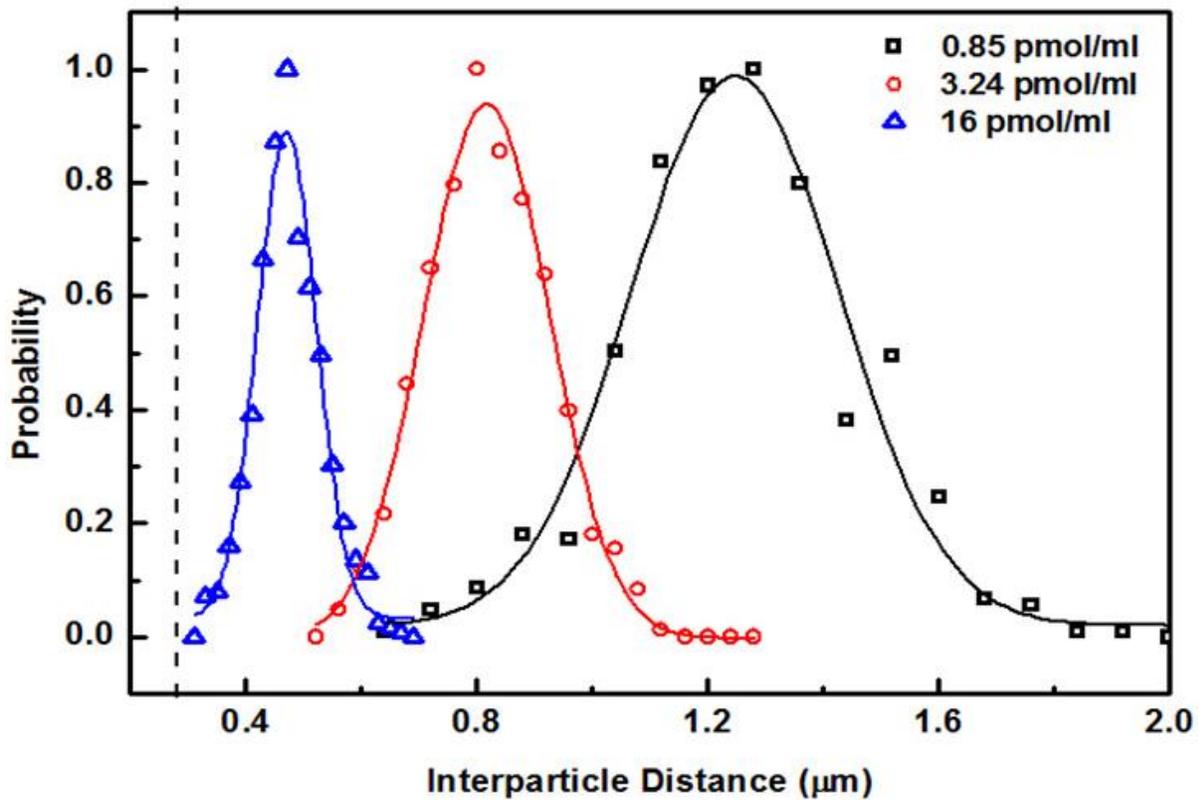

**Figure 5.** Probability distribution of the inter-particle spacing of UCNPs dispersed in nematic LC, for concentrations 0.85 pmol/ml (□), 3.24 pmol/ml (○) and 16 pmol/ml (Δ). Solid lines represent the Gaussian fits to the experimentally measured distributions. The vertical dashed line indicates the theoretical diffraction limit of the optical microscope.